\documentclass[english]{article}
\usepackage[T1]{fontenc}
\usepackage[latin1]{inputenc}
\usepackage{babel}

\makeatletter

\providecommand{\LyX}{L\kern-.1667em\lower.25em\hbox{Y}\kern-.125emX\@}

\usepackage{amsfonts}

\makeatother
\begin{document}
\begin{flushright}

MPP-2003-138

\end{flushright}

{\centering \textbf{\LARGE NC Wilson lines and the inverse Seiberg-Witten
map for nondegenerate star products}\LARGE \par}

\textbf{\bigskip}

\bigskip

{\centering Wolfgang Behr%
\footnote{behr@theorie.physik.uni-muenchen.de
}, Andreas Sykora%
\footnote{andreas@theorie.physik.uni-muenchen.de
}\par}

\bigskip

{\centering Max-Planck-Institut f\"ur Physik, \par}

{\centering F\"ohringer Ring 6, 80805 M\"unchen, Germany\par}

\begin{abstract}
Open Wilson lines are known to be the observables of noncommutative
gauge theory with Moyal-Weyl \( \star  \)-product. We generalize
these objects to more general \( \star  \)-products. As an application
we derive a formula for the inverse Seiberg-Witten map for \( \star  \)-products
with invertible Poisson structures. 
\end{abstract}

\newpage

\subsection*{Introduction}

Noncommutative gauge theories have been under strong investigation
as was realized that they can be implemented by a certain string theory
\cite{Seiberg:1999vs}. In these theories the noncommutativity is
introduced via a \( \star  \)-product on ordinary function spaces.
Most research up to now has only considered the case of the Moyal-Weyl
\( \star  \)-product \[
f\star g=\lim _{x^{\prime }\rightarrow x}\, e^{\frac{i}{2}\theta ^{ij}\partial _{i}\partial ^{\prime }_{j}}f(x)(x^{\prime })\]
that depends on a constant tensor \( \theta ^{ij} \). In the string
theory approach, this tensor is related to a constant \( B \)-field
on a brane. On a curved brane this \( B \)-field becomes position
dependent \cite{Cornalba:2001sm}. For this it is necessary to look
at the case where the tensor \( \theta ^{ij} \) is not any more constant.
In this case \( \star  \)-products can be defined as polydifferential
operators \begin{equation}
\label{DEF_STAR_PRODUCT}
f\star g=fg+\frac{i}{2}\theta ^{ij}(x)\partial _{i}f\partial _{j}g+\mbox {higher\, order\, terms}.
\end{equation}
 The functions should still form an associative algebra. Therefore
\( \theta ^{ij} \) has to be a Poisson tensor\[
\theta ^{il}\partial _{l}\theta ^{jk}+cyc.=0.\]
On the other hand one can show that every Poisson tensor gives rise
to a \( \star  \)-product that looks like (\ref{DEF_STAR_PRODUCT})
\cite{Kontsevich:1997vb} and that a large class of algebras may be
represented in this way \cite{Behr:2003qc}. In these cases the higher
order terms all depend only on \( \theta ^{ij} \) and its derivatives.

The fundamental objects of noncommutative gauge theory are covariant
coordinates which can also be defined for the \( \star  \)-products
(\ref{DEF_STAR_PRODUCT}). In this paper we will use these covariant
coordinates to generalize the open Wilson lines introduced in \cite{Ishibashi:1999hs}.
In \cite{Okawa:2001mv} they were used to give an exact formula for
the inverse Seiberg-Witten map. We will generalize this construction
for \( \star  \)-products of type (\ref{DEF_STAR_PRODUCT}) with
invertible Poisson structure \( \theta ^{ij} \).

\subsection*{Covariant coordinates}

In a noncommutative version of a \( U(1) \)-gauge theory, a scalar
field should transform like\[
\phi ^{\prime }=g\star \phi ,\]
 where \( g \) is a function that is invertible with respect to the
\( \star  \)-product\[
g\star g^{-1}=g^{-1}\star g=1.\]
Note that multiplication with a coordinate function is not covariant
any more\[
(x^{i}\star \phi )^{\prime }\neq x^{i}\star \phi ^{\prime }.\]
In the classical case the same problem arises with the partial derivatives.
In analogy to this, covariant coordinates\[
X^{i}(x)=x^{i}+A^{i}(x)\]
 can be introduced transforming in the adjoint representation\[
X^{i\prime }=g\star X^{i}\star g^{-1}.\]
Now the product of a covariant coordinate with a field is again a
field. An infinitesimal version of this is presented in \cite{Madore:2000en}.
The equivalence of both approaches is investigated in \cite{Jurco:2001rq}. 

In \cite{Seiberg:1999vs} it was shown that commutative and noncommutative
gauge theory can be related by the so called Seiberg-Witten-map. Mapping
the classical gauge transformations and gauge fields to their noncommutative
counterparts, one can show that \[
A^{i}=\theta ^{ij}a_{j}+\mathcal{O}(\theta ^{2}).\]
This equality also holds in the case of the \( \star  \)-products
(\ref{DEF_STAR_PRODUCT}) (see \cite{Jurco:2000fs}).

\subsection*{Wilson lines}

In the case \( \theta ^{ij}=const. \) the basic observation was that
translations in space are gauge transformations. \cite{Ishibashi:1999hs}
They are realized by \[
T_{l}x^{j}=x^{j}+l_{i}\theta ^{ij}=e^{il_{i}x^{i}}\star f\star e^{-il_{i}x^{i}}.\]
Now one can pose the question what happens if one uses covariant coordinates.
In this case the inner automorphism\[
f\rightarrow e^{il_{i}X^{i}}\star f\star e^{-il_{i}X^{i}}\]
should consist of a translation and a gauge transformation dependent
of the translation. If we subtract the translation again only the
gauge transformation remains and the resulting object \[
W_{l}=e^{il_{i}X^{i}}\star e^{-il_{i}x^{i}}\]
has a very interesting transformation behavior under a gauge transformation\[
W_{l}^{\prime }(x)=g(x)\star W_{l}(x)\star g^{-1}(x+l_{i}\theta ^{ij}).\]
It transforms like a Wilson line starting at \( x \) and ending at
\( x+l\theta  \).

As in the constant case we can start with \[
W_{l}=e^{il_{i}X^{i}}_{\star }\star e^{-il_{i}x^{i}}_{\star },\]
where now \( e_{\star } \) is the \( \star  \)-exponential. Every
multiplication in its Taylor series is replaced by the \( \star  \)-product.
In contrast to the constant case, \( e^{f}_{\star }=e^{f} \) isn't
true any more true. The transformation property of \( W_{l} \) is
now\[
W^{\prime }_{l}(x)=g(x)\star W_{l}(x)\star g^{-1}(T_{l}x),\]
where \[
T_{l}x^{j}=e^{il_{i}x^{i}}\star x^{j}\star e^{-il_{i}x^{i}}\]
is an inner automorphism of the algebra, which can be interpreted
as a quantized coordinate transformation. If we replace commutators
by Poisson brackets the classical limit of this coordinate transformations
may be calculated\[
T_{l}x^{k}=e^{il_{i}[x^{i}\stackrel{\star }{,}\cdot ]}x^{k}\approx e^{-l_{i}\{x^{i},\cdot \}}x^{j}=e^{-l_{i}\theta ^{ij}\partial _{j}}x^{k},\]
the formula becoming exact for \( \theta ^{ij} \) constant or linear
in \( x \). We see that the classical coordinate transformation is
the flow induced by the Hamiltonian vector field \( -l_{i}\theta ^{ij}\partial _{j} \).
At the end we may expand \( W_{l} \) in terms of \( \theta  \) and
get\[
W_{l}=e^{il_{i}\theta ^{ij}a_{j}}+\mathcal{O}(\theta ^{2}),\]
where we have replaced \( A^{i} \) by its Seiberg-Witten expansion.
We see that for \( l \) small this really is a Wilson line starting
at \( x \) and ending at \( x+l\theta  \).

\subsection*{Observables}

As space translations are included in the noncommutative gauge transformations
no local observables can be constructed. In the case \( \theta ^{ij}=const. \)
one has to integrate over the whole space\[
U_{l}=\int d^{n}x\, W_{l}(x)\star e^{il_{i}x^{i}}.\]
If one goes to the Fock space representation of the algebra one sees
that this corresponds to\[
U_{l}=tr\, e^{il_{i}\hat{X}^{i}}.\]
In the more general case of non constant \( \theta ^{ij} \) we therefore
need a trace for the \( \star  \)-product, i. e. a functional \( tr \)
with the property \[
tr\, f\star g=tr\, g\star f.\]
Only in this case the trace of a covariant quantity becomes invariant.
For many \( \star  \)-products we know that the trace may be written
as\begin{equation}
\label{DEF_OMEGA}
tr\, f=\int d^{n}x\, \Omega (x)\, f(x)
\end{equation}
with a measure function \( \Omega  \). Due to the cyclicity of the
trace it has to fulfill\begin{equation}
\label{PROP_OMEGA}
\partial _{i}(\Omega \theta ^{ij})=0
\end{equation}
which can easily be calculated with (\ref{DEF_STAR_PRODUCT}). If
the Poisson structure \( \theta ^{ij} \) is invertible then the inverse
of the Pfaffian\[
\frac{1}{\Omega }=Pf(\theta )=\sqrt{det(\theta )}=\frac{1}{2^{n}n!}\epsilon _{i_{1}i_{2}\cdots i_{2n}}\theta ^{i_{1}i_{2}}\cdots \theta ^{i_{2n-1}i_{2n}}\]
is a solution to this equation. There can always be found a \( \star  \)-product
so that a measure function fulfilling (\ref{PROP_OMEGA}) guarantees
cyclicity to all orders \cite{Felder:2000hy}. Now we are able to
write down a large class of observables for the above defined noncommutative
gauge theory, namely\[
U_{l}=\int d^{n}x\, \Omega (x)\, W_{l}(x)\star e_{\star }^{il_{i}x^{i}}=\int d^{n}x\, \Omega (x)\, e^{il_{i}X^{i}(x)}_{\star }\]
or more general\[
f_{l}=\int d^{n}x\, \Omega (x)\, f(X^{i})\star e^{il_{i}X^{i}(x)}_{\star }\]
with \( f \) an arbitrary function of the covariant coordinates.

\subsection*{Inverse Seiberg-Witten-map}

As an application of the above constructed observables we generalize
\cite{Okawa:2001mv} to arbitrary \( \star  \)-products, i. e. we
give a formula for the inverse Seiberg-Witten map for \( \star  \)-products
with invertible Poisson structure. In order to map noncommutative
gauge theory to its commutative counterpart we need a functional \( f_{ij}[X] \)
fulfilling 

\[
f_{ij}[g\star X\star g^{-1}]=f_{ij}[X],\]
\[
df=0\]
and\[
f_{ij}=\partial _{i}a_{j}-\partial _{j}a_{i}+\mathcal{O}(\theta ^{2}).\]
 \( f \) is a classical field strength and reduces in the limit \( \theta \rightarrow 0 \)
to the correct expression.

To prove the first and the second property we will only use the algebra
properties of the \( \star  \)-product and the cyclicity of the trace.
All quantities with a hat will be elements of an algebra. With this
let \( \hat{X}^{i} \) be covariant coordinates in an algebra, transforming
under gauge transformations like\[
\hat{X}^{i\prime }=\hat{g}\hat{X}^{i}\hat{g}^{-1}\]
with \( \hat{g} \) an invertible element of the algebra. Now define
\[
\hat{F}^{ij}=-i[\hat{X}^{i},\hat{X}^{j}]\]
and \[
(\hat{F}^{n-1})_{ij}=\frac{1}{2^{n-1}(n-1)!}\epsilon _{iji_{1}i_{2}\cdots i_{2n-2}}\hat{F}^{i_{1}i_{2}}\cdots \hat{F}^{i_{2n-3}i_{2n-2}}.\]
Since an antisymmetric matrix in odd dimensions is never invertible
we have assumed that the space is \( 2n \) dimensional. The expression\begin{equation}
\label{DEF_F}
\mathcal{F}_{ij}(k)=str_{\hat{F},\hat{X}}\left( (\hat{F}^{n-1})_{ij}e^{ik_{j}\hat{X}^{j}}\right) 
\end{equation}
clearly fulfills the first property due to the properties of the trace.
\( str \) is the symmetrized trace, for an exact definition see \cite{Okawa:2001mv}.
Note that symmetrization is only necessary for space dimension bigger
than 4 due to the cyclicity of the trace. In dimensions 2 and 4 we
may replace \( str \) by the ordinary trace \( tr \). \( \mathcal{F}_{ij}(k) \)
is the Fourier transform of a closed form if \[
k_{[i}\mathcal{F}_{jk]}=0\]
or if the current \[
J^{i_{1}\cdots i_{2n-2}}=str_{\hat{F},X}\left( \hat{F}^{[i_{1}i_{2}}\cdots \hat{F}^{i_{2n-3}i_{2n-2}]}e^{ik_{j}\hat{F}^{j}}\right) \]
is conserved, respectively\[
k_{i}J^{i\cdots }=0.\]
This is easy to show, if one uses\[
str_{\hat{F},\hat{X}}\left( [k\hat{X},\hat{X}^{l}]e^{ik_{j}\hat{X}^{j}}\cdots \right) =str_{\hat{F},\hat{X}}\left( [\hat{X}^{l},e^{ik_{j}\hat{X}^{j}}]\cdots \right) =str_{\hat{F},\hat{X}}\left( e^{ik_{j}\hat{X}^{j}}[\hat{X}^{l},\cdots ]\right) \]
which can be calculated by simple algebra.

To prove the last property we have to switch to the \( \star  \)-product
formalism and expand the formula in \( \theta ^{ij} \). The expression
(\ref{DEF_F}) now becomes \[
\mathcal{F}[X]_{ij}(k)=\int \frac{d^{2n}x}{Pf(\theta )}\left( (F_{\star }^{n-1})_{ij}\star e_{\star }^{ik_{j}X^{j}}\right) _{sym\, F,X}.\]
The expression in brackets has to be symmetrized in \( F^{ij} \)
and \( X^{i} \) for \( n>2 \). Up to third order in \( \theta ^{ij} \),
the commutator \( F^{ij} \) of two covariant coordinates is \[
F^{ij}=-i[X^{i}\stackrel{\star }{,}X^{j}]=\theta ^{ij}-\theta ^{ik}f_{kl}\theta ^{lj}-\theta ^{kl}\partial _{l}\theta ^{ij}a_{k}+\mathcal{O}(3)\]
with \( f_{ij}=\partial _{i}a_{j}-\partial _{j}a_{i} \) the ordinary
field strength. Furthermore we have\[
e^{ik_{i}X^{i}}_{\star }=e^{ik_{i}x^{i}}(1+ik_{i}\theta ^{ij}a_{j})+\mathcal{O}(2).\]
If we choose the antisymmetric \( \star  \)-product (\ref{DEF_STAR_PRODUCT}),
the symmetrization will annihilate all the first order terms of the
\( \star  \)-products between the \( F^{ij} \) and \( X^{i} \),
and therefore we get

\begin{eqnarray*}
 &  & -\mathcal{F}[X]_{ij}(k)\\
 &  & =-2n\int \frac{d^{2n}x}{\epsilon \theta ^{n}}\left( \epsilon _{ij}\theta ^{n-1}-(n-1)\epsilon _{ij}\theta ^{n-2}\theta f\theta -\theta ^{kl}\partial _{l}(\epsilon _{ij}\theta ^{n-1})a_{k}\right) e^{ik_{i}x^{i}}+\mathcal{O}(1)\\
 &  & =-2n\int \frac{d^{2n}x}{\epsilon \theta ^{n}}\left( \epsilon _{ij}\theta ^{n-1}-(n-1)\epsilon _{ij}\theta ^{n-2}\theta f\theta -\frac{1}{2}\epsilon _{ij}\theta ^{n-1}f_{kl}\theta ^{kl}\right) e^{ik_{i}x^{i}}+\mathcal{O}(1)\\
 &  & =d^{2n}x\, \left( \theta _{ij}^{-1}+2n(n-1)\frac{\epsilon _{ij}\theta ^{n-2}\theta f\theta }{\epsilon \theta ^{n}}-\frac{1}{2}\theta _{ij}^{-1}f_{kl}\theta ^{kl}\right) e^{ik_{i}x^{i}}+\mathcal{O}(1),
\end{eqnarray*}
using partial integration and \( \partial _{i}(\epsilon \theta ^{n}\theta ^{ij})=0 \).
To simplify notation we introduced \( \epsilon _{ij}\theta ^{n-1}=\epsilon _{iji_{1}j_{1}\cdots i_{n-1}j_{n-1}}\theta ^{i_{1}j_{1}}\cdots \theta ^{i_{n-1}j_{n-1}} \)
etc. In the last line we have used\[
\theta _{ij}^{-1}=-\frac{(\theta ^{n-1})_{ij}}{Pf(\theta )}=-2n\frac{\epsilon _{ij}\theta ^{n-1}}{\epsilon \theta ^{n}}.\]
We will now have a closer look at the second term, noting that\[
\theta ^{ij}\frac{\epsilon _{ij}\theta ^{n-2}\theta f\theta }{\epsilon \theta ^{n}}=-\frac{1}{2n}\theta ^{-1}_{kl}\theta ^{kr}f_{rs}\theta ^{sl}=-\frac{1}{2n}f_{rs}\theta ^{rs}\]
and therefore\begin{equation}
\label{defining formula for a b}
\frac{\epsilon _{ij}\theta ^{n-2}\theta f\theta }{\epsilon \theta ^{n}}=a\frac{\epsilon _{ij}\theta ^{n-1}}{\epsilon \theta ^{n}}f_{rs}\theta ^{rs}+bf_{ij}
\end{equation}
with \( a+b=-\frac{1}{2n} \). Taking e. g. \( i=1,j=2 \) we see
that\[
\epsilon _{12\cdots kl}\theta ^{n-2}\theta ^{kr}f_{rs}\theta ^{sl}=\epsilon _{12\cdots kl}\theta ^{n-2}(\theta ^{k1}\theta ^{2l}-\theta ^{k2}\theta ^{1l})f_{12}+\mbox {terms\, without}\: f_{12}.\]
 Especially there are no terms involving \( f_{12}\theta ^{12} \)
and we get for the two terms on the right hand side of (\ref{defining formula for a b})\[
2a\epsilon _{12}\theta ^{n-1}f_{12}\theta ^{12}=-2nb\epsilon _{12}\theta ^{12}\theta ^{n-1}f_{12}\]
 and therefore \( b=-\frac{a}{n} \). This has the solution \[
a=-\frac{1}{2(n-1)}\: \: \: \mbox {and}\: \: \: b=\frac{1}{2n(n-1)}.\]
With the resulting 

\[
2n(n-1)\frac{\epsilon _{ij}\theta ^{n-2}\theta f\theta }{\epsilon \theta ^{n}}=\frac{1}{2}\theta _{ij}^{-1}f_{kl}\theta ^{kl}+f_{ij}\]
we finally get \[
-\mathcal{F}[X]_{ij}(k)=\int d^{2n}x\, \left( \theta _{ij}^{-1}+f_{ij}\right) e^{ik_{i}x^{i}}+\mathcal{O}(1).\]
Therefore\[
f[X]_{ij}=\mathcal{F}[X]_{ij}(k)-\mathcal{F}[x]_{ij}(k)\]
is a closed form that reduces in the classical limit to the classical
Abelian field strength. We have found an expression for the inverse
Seiberg-Witten map.

\subsection*{Outlook}

It would be interesting to find a similar expression to (\ref{DEF_F})
for other noncommutative (compact) spaces like the fuzzy torus and
the fuzzy sphere. In the second case we would be able to map an commutative
\( su(2) \)-gauge theory to an commutative Abelian gauge theory in
higher dimensions.

\subsection*{Acknowledgements}

The authors want to thank B. Jurco for helpful discussions. Also we
want to thank the MPI and the LMU for their support.

\bibliographystyle{diss}
\bibliography{mainbib}

\end{document}